\documentclass[
reprint,
superscriptaddress,
 amsmath,amssymb,
aps,
prb,
]{revtex4-2}

\usepackage{amsmath}
\usepackage{graphicx}
\usepackage{dcolumn}
\usepackage{hyperref}
\usepackage{textcomp}
\usepackage[dvipsnames]{xcolor}
\usepackage{bm}
\usepackage{hyperref}
\usepackage{color} 
\usepackage{soul} 
\usepackage{mathtools}
\usepackage{tikz}
\newcommand{\reviewing}[1]{{\leavevmode\color{black}{#1}}}
\newcommand*\circled[1]{\tikz[baseline=(char.base)]{
            \node[shape=circle,draw,inner sep=1pt] (char) {#1};}}

\begin{document}

\preprint{APS/123-QED}

\title{Sampling from exponential distributions in the time domain with superparamagnetic tunnel junctions }
\author{Temitayo N. Adeyeye}
\author{Sidra Gibeault}
\affiliation{%
Institute for Research in Electronics and Applied Physics, University of Maryland, College Park, MD, USA
}%
\affiliation{
Associate, Physical Measurement Laboratory, National Institute of Standards and Technology, Gaithersburg, MD, USA
}%

\author{Daniel P. Lathrop}%
\affiliation{%
Institute for Research in Electronics and Applied Physics, University of Maryland, College Park, MD, USA
}%


\author{Matthew W. Daniels}
\author{Mark D. Stiles}
\author{Jabez J. McClelland}
\author{William A. Borders}
\author{Jason T. Ryan}
\affiliation{
Physical Measurement Laboratory, National Institute of Standards and Technology, Gaithersburg, MD, USA
}%
\author{Philippe Talatchian}
\author{Ursula Ebels}
\affiliation{%
Univ.~Grenoble Alpes, CEA, CNRS, Grenoble INP, SPINTEC, 38000 Grenoble, France
}%

\author{Advait Madhavan}
\affiliation{%
Institute for Research in Electronics and Applied Physics, University of Maryland, College Park, MD, USA
}%
\affiliation{
Associate, Physical Measurement Laboratory, National Institute of Standards and Technology, Gaithersburg, MD, USA
}%


\begin{abstract}

\reviewing{In the superparamagnetic regime, magnetic tunnel junctions switch between two resistance states due to random thermal fluctuations. The dwell time distribution in each state is exponential. We sample this distribution using a temporal encoding scheme, in which information is encoded in the time at which the device switches between its resistance states.} We then develop a circuit element known as a probabilistic delay cell that applies an electrical current step to a superparamagnetic tunnel junction and a temporal measurement circuit that measures the timing of the first switching event. Repeated experiments confirm that these times are exponentially distributed. Temporal processing methods then allow us to digitally compute with these exponentially distributed probabilistic delay cells. We describe how to use these circuits in a Metropolis-Hastings stepper and in a weighted random sampler, both of which are computationally intensive applications that benefit from the efficient generation of exponentially distributed random numbers.

\end{abstract}

\maketitle




\section{\label{sec:intro}Introduction}








\reviewing{Though exponential distributions are ubiquitous in statistical physics and related computational models, directly sampling them from device behavior is rarely done. The magnetic tunnel junction (MTJ) operating in the superparamagnetic regime, a key device in probabilistic computing, is known to naturally exhibit exponentially distributed temporal switching dynamics~\cite{vodenicarevic2017low,hayakawa2021nanosecond,talatchian2021mutual}. Contemporary approaches focus on sampling the instantaneous state of the device~\cite{niazi2024training,borders2019integer} and treat each switching event as a Bernoulli trial resulting in a discrete sample. To sample an exponential distribution with an MTJ, we need to measure it in the time domain, which is challenging with traditional techniques. A time-domain-based sampling technique would result in exponentially distributed analog time values, which can be further processed based on application requirements. }

The importance of the exponential distribution cannot be overstated. Characterized by a rate parameter, it is central to any Markovian description of evolution through state space and to any energy-based objective function description since the Boltzmann distribution that describes the occupancy of the states at equilibrium is an exponential function of their energies~\cite{kittel1980thermal}. This property is exploited in the Metropolis-Hastings algorithm for Markov chain Monte Carlo sampling, by accepting transitions to higher energy states with a probability proportional to an exponential function of the difference in energies~\cite{kirkpatrick1983optimization}. Exponential distributions can also be used to generate weighted random samples by selecting the argument of the smallest drawn sample from a set of exponential variates, a technique known as the ``exponential clocks method"~\cite{hubschle2022parallel}. This can be used to generate arbitrarily weighted random samples, akin to rolling an arbitrarily biased $n$-sided die. 

While sampling techniques based on measuring the instantaneous states of MTJ dynamics are extremely useful for probabilistic computations, using the resulting Bernoulli distributions to sample from exponential distributions would require significant memory and computational overhead. The exponential distributions reside in the temporal characteristics, not state distributions, of SMTJ switching events. The ability to efficiently sample exponential distributions has the potential to enhance the efficiency of different types of probabilistic computing.

\begin{figure}
    \includegraphics[width=8.5cm]{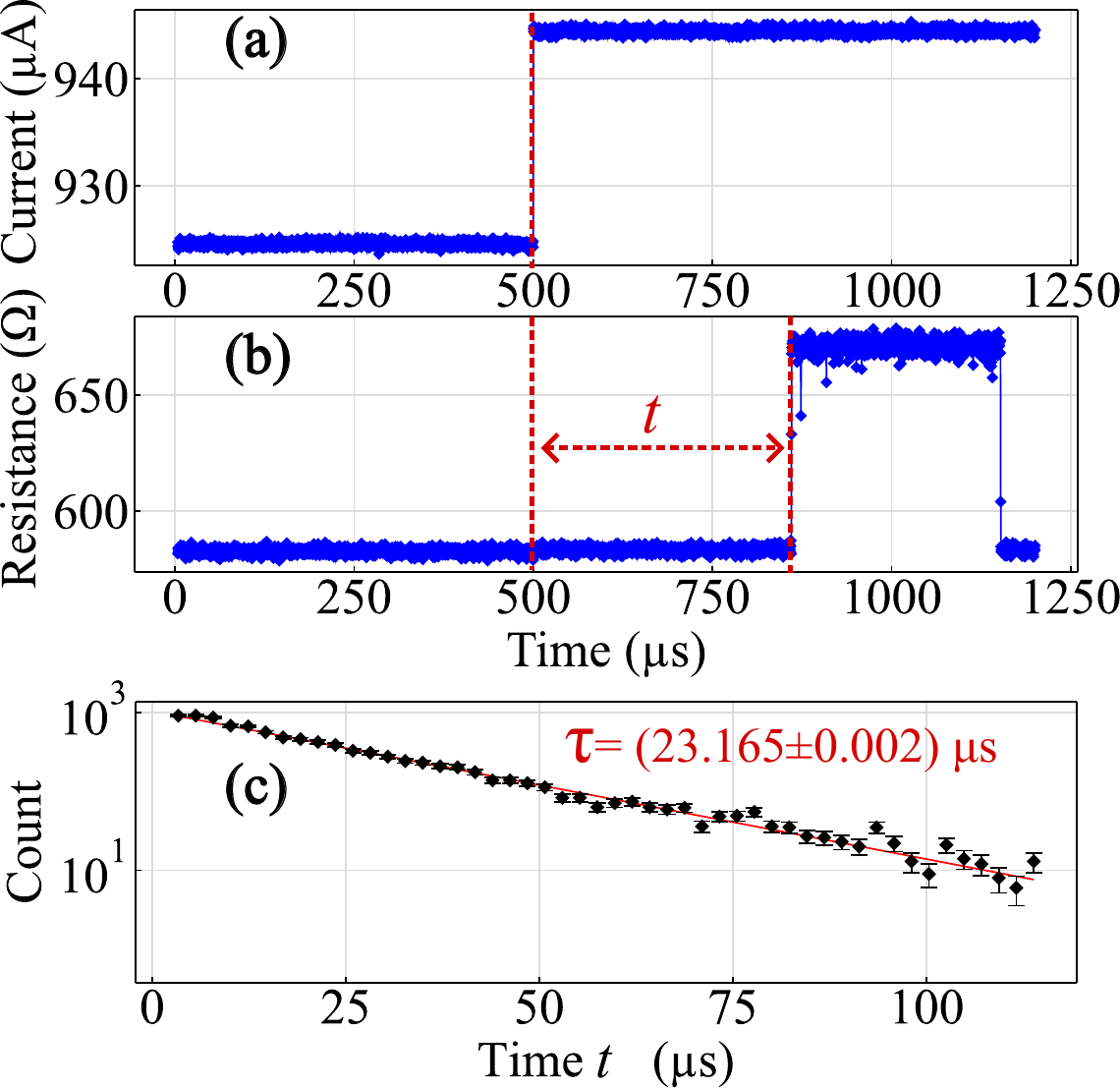}
    \caption{Extracting temporal stochasticity with a current step: Panel (a) shows the current flowing through the device as a function of time. Initially, a low current value keeps the device in the parallel state. At a clock edge, a current step is applied across the device, putting it in the superparamagnetic state,  causing a switching event after some delay as is shown in panel (b). The time taken for this switching event $t$ is the quantity of interest, and is measured using on-board electronics. Panel (c) shows the experimentally measured exponential distribution of the times of the first arriving switch, as measured with the circuit shown in Fig.~\ref{fig:ckt-diagram}(a). The slope of the straight line fit to the data determines the empirical mean $\tau = \langle t\rangle$. The error bars are calculated as the square root of the number of entries in each bin. Approximately 10,000 switching events were collected for a current step with a constant amplitude and width. While the statistical uncertainties are smaller at longer times due to the reduced number of counts, they appear larger on the logarithmic scale. The reduced chi-squared statistic for the fit is $\chi^2$ = 1.92, indicating that the statistical variations provide the primary source of uncertainty in the data. }
    \label{fig:intro}
\end{figure}

In this work, we design a circuit aimed at drawing samples from exponential distributions by measuring the temporal statistics of the first switching events of MTJs when exposed to a current step. As shown in Fig.~\ref{fig:intro}(a,b), the MTJ is first biased to be in the parallel state just outside the superparamagnetic regime (on experimentally relevant time scales). Subsequently, a current step is applied, putting the MTJ in a superparamagnetic regime where there is eventually a transition to the antiparallel state after a probabilistic delay that depends on the magnitude of the current step. The time elapsed before this switching event occurs (shown in Fig.~\ref{fig:intro}(c)) is measured with embedded electronics and is then used to demonstrate the exponential switching characteristics of this probabilistic delay cell (PDC). By creating switching events that are temporally encoded and drawn from an exponential distribution, we can utilize this delay cell alongside other temporal computing primitives to make Boolean decisions based on the arrival times of both deterministic and stochastic temporal signals. We extend this approach to the ``exponential clocks method" and propose a circuit to draw weighted random samples from arbitrary distributions.

The rest of the paper is organized as follows. Section~\ref{sec:background} gives background information on probabilistic computing, magnetic tunnel junctions, and come existing methods of producing random samples. Section~\ref{sec:CDES} describes the circuit and the experimental setup used to interrogate the device and collect data. Section~\ref{sec:delay_char} describes the statistical properties of the first switching event and its tunability with current. Section~\ref{sec:appliction} describes how these circuits can be used in applications such as a Metropolis-Hastings stepper and a weighted random sampler, followed by a discussion in Sec.~\ref{sec:discussion}.

\section{\label{sec:background}Background}

Probabilistic computing~\cite{chowdhury2023full,misra2023probabilistic} is a modern computing paradigm inspired by the statistical physics of interacting bodies, such as gas molecules in a box~\cite{camsari2024probabilistic}, metal atoms during an annealing process~\cite{kirkpatrick1983optimization}, or magnetic moments in a ferromagnetic material~\cite{cipra1987introduction}. Various combinatorial optimization problems with real-world utility can be solved by formulating the constraints of the problem into an objective function, which is analogous to the energy landscape of a physical system~\cite{kirkpatrick1983optimization}. The objective function mathematically describes the interactions between participating bodies in configuration space. The configuration that minimizes this objective function corresponds to the solution~\cite{kirkpatrick1983optimization}. \reviewing{Broadly speaking, there are two approaches to probabilistic computing. One approach~\cite{camsari2017stochastic,gibeault2024programmable} encodes the configuration space of such problems in probabilistic bits, also known as $p$-bits, each of which can take one of two binary values, 0 or 1, but do so stochastically with probabilities that depend on their interactions with their neighbors. These probabilistic bits interact through carefully engineered electrical interactions, which implement the objective function to be minimized. The other approach, which is more relevant to the present manuscript, stores the state of the system in deterministic bits and uses a physically separate random process to determine switching dynamics.} 

The generality of being able to write objective functions for a wide variety of problems such as prime factorization~\cite{borders2019integer}, reversible Boolean gates~\cite{camsari2017stochastic}, circuit satisfiability~\cite{aadit2022massively}, probabilistic inference~\cite{singh2024cmos}, machine learning with Boltzmann machines~\cite{niazi2024training}, traveling salesman problems~\cite{si2023energy}, optimal control~\cite{hespanha2024markov}, and many others~\cite{lucas2014ising} indicates the broad scope and the potential usefulness of probabilistic computers. The performance of such systems depends on the performance of their functional blocks that sample random processes appropriate for application-constrained distributions. Though modern computing systems possess these functional units, such as on-chip physical random number generators~\cite{hamburg2012analysis,NVIDIATRNG} and Boolean computing units, they cannot be efficiently implemented at the scale and granularity required to perform meaningful probabilistic computations for real-world applications. Some of these applications require upwards of $10^{12}$ random bits per second~\cite{niazi2024training}. 

To address these challenges, recent work in the field has relied on novel device technologies to sample large quantities of random bits from device-specific distributions while using the reconfigurability and granularity of modern field programmable gate arrays (FPGAs) to implement the interaction circuits~\cite{niazi2024training,singh2024cmos,aadit2022massively}. Though this approach is suitable for exploring application space, the ultimate goal of orders of magnitude increases in computing performance will require integrated implementations. Integrated hardware accelerators of this kind require careful co-design between various levels of the computing stack such as integration of novel device technologies in the back-end-of-the-line, efficient circuits for sampling random bits, and mapping of problems onto scalable architectures~\cite{wan2022compute}. Fortunately, one novel device technology---the superparamagnetic magnetic tunnel junction, 
(SMTJ)~\cite{vodenicarevic2017low}, which has been widely
used for probabilistic computing~\cite{aadit2021computing,camsari2017stochastic,borders2019integer}---is a low-barrier version of a mature, scalable, and integrable memory technology and has the potential to meet all of the aforementioned needs.

A simplified description of the magnetic tunnel junction (MTJ) device stack consists of two ferromagnetic layers separated by an insulating tunneling barrier. The resistance of the stack depends on the relative orientations of the ferromagnetic layer magnetizations: parallel magnetizations (P) have lower resistance, and antiparallel magnetizations (AP) have higher resistance~\cite{stiles2002anatomy}. The resistance state can be measured with a small read-current and can be switched via an applied magnetic field or a larger write-current. For long-term storage ($\approx 10$ years), the energy barrier between states needs to be $\geq 40~kT$, where $k$ is the Boltzmann constant and $T =300$~K is room temperature~\cite{lee20191gbit,worledge2022spin}. \reviewing{When the device is biased to favor a different state than it currently exhibits, it undergoes a stochastic switching process, with the time scale of the switching controlled by the size of the bias \cite{zhao2011sub}.

If the barrier between the states is reduced, the retention time in both states drops exponentially, and for low enough barriers the device exhibits random telegraph fluctuations at the millisecond or smaller time scales at room temperature~\cite{vodenicarevic2017low,gibeault2024programmable,schnitzspan2023nanosecond,hayakawa2021nanosecond}.} In this regime, the device is said to be in a superparamagnetic state. Small currents applied across an SMTJ can both probe the stochastic device state as a voltage and also tune the mean time the device spends in each state. Both of these properties are exploited in implementations of $p$-bit circuits~\cite{gibeault2024programmable,borders2019integer,si2023energy}. \reviewing{Particular devices may have fringing fields acting on them that shift the superparamagnetic regime to be centered around a finite bias.}

Previous proposals and demonstrations of probabilistic computing systems measure the instantaneous state of the SMTJ state as a source of random bits. Even though the measured voltage itself is an analog quantity that exhibits a bimodal distribution~\cite{talatchian2021mutual}, the thresholded digital output voltages are very good approximations of samples from a Bernoulli distribution with probability $p$ of getting a true outcome. By changing the current flowing through the SMTJ, the value of $p$ can be varied. Most commonly, SMTJs are biased to their balanced state, where the probability $p \approx 0.5$~\cite{vodenicarevic2017low}. Most recent FPGA~\cite{niazi2024training,singh2024cmos,aadit2022massively} and microprocessor-based~\cite{si2023energy} implementations use digitized versions of such SMTJs as entropy sources to seed probabilistic computations. 

\reviewing{One of the first high-barrier MTJ-state-based random number generator designs was the ``spin dice'' approach~\cite{Fukushima_2014}, This approach utilizes a precisely tuned current pulse to bring an MTJ to a bifurcation point and then allows the system to transition into one of its stable magnetic states. Although the MTJ features well-defined, stable magnetic configurations separated by an energy barrier, thermal fluctuations at the bifurcation point affect the magnetization dynamics, leading to an equal probability of settling into either final state when the pulse is tuned correctly. In such proposals, the energy dissipation can be reduced by using medium barrier magnets~\cite{shukla2023}, which can provide functionally similar operations as more traditional high barrier magnets with lower power dissipation.

Further reducing the barrier causes the MTJ to operate in a superparamagnetic state, exhibiting random telegraph switching behavior due to the amplification of inherent thermal noise. By relying on thermal fluctuations for switching, this approach eliminates the need for external triggering or resetting pulses, thereby reducing energy consumption. However, reading out the stochastic fluctuations still necessitates a small, continuously applied sense current~\cite{camsari2017stochastic}. Circuit techniques can increase energy efficiency~\cite{vodenicarevic2017low} by capturing these fluctuations without the need for a static current.

These approaches to random bit generation generate, at most, a single Bernoulli trial, which is generally at least slightly biased. In theory, each trial can generate bit of randomness per switching event, but this is difficult to achieve in practice and requires tight fabrication margins. A solution to this problem can be traced back to von Neumann~\cite{von1963various}, who suggested that in a physical system it is much easier to achieve independence between two events than it is to obtain a perfectly unbiased distribution on each trial. He proposed to obtain unbiased outcomes by flipping two biased coins and selecting certain joint outcomes while discarding others. The heads-heads and tails-tails tosses are discarded since they have unequal probabilities, while the heads-tails and tails-heads tosses are kept since both have the same probability $p(1-p)$. Treating each of the latter configurations as a single statistical event allows for two of the four samples from a Bernoulli generator to accurately generate an unbiased distribution. This approach has been generalized to a tree-based data structure called a Bernoulli factory. 

A Bernoulli factory can be thought of as a binary tree constructed out of a sequence of coin-tosses, with a biased coin that has a probability $p$ of being heads~\cite{keane1994bernoulli}. Each node in the tree has two branches, with one branch representing heads and the other representing tails. Bernoulli factories can be either finite or infinite and are used to sample from discrete distributions as a function of $p$. Many algorithmic techniques for generating uniform random bits, weighted random samples, and exponential distributions can be formulated as Bernoulli factories~\cite{dughmi2021bernoulli,keane1994bernoulli}. All of the techniques require repeated Bernoulli trials, which are conventionally implemented with a state-encoded approach to random number generation. This has resulted in a race towards building faster and more efficient SMTJ devices~\cite{hayakawa2021nanosecond,schnitzspan2023nanosecond,soumah2024nanosecond}. Though tunability allows adjusting the value of $p$ that the device naturally produces, it doesn't fundamentally change the discrete nature of such a Bernoulli event. The next section shows how the ~\emph{temporal domain is qualitatively different since it allows an analog quantity to be sampled from a single SMTJ switching event.}}


\begin{figure*}[ht]
    \includegraphics[width=\textwidth]{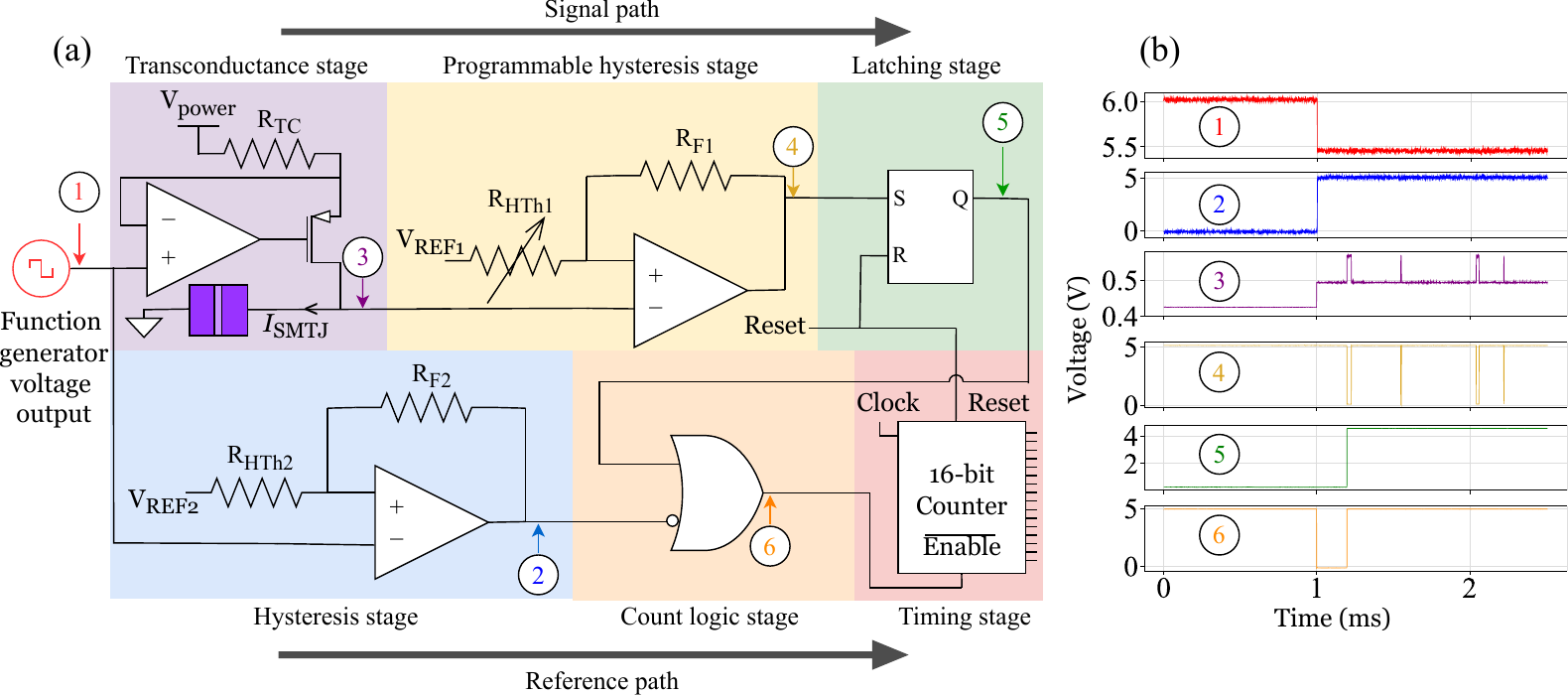}
    \caption{Onboard timing measurement of a probabilistic delay cell: Panel (a) shows the timing measurement circuit with the signal and reference path culminating in the clocked counter. The stages that comprise the probabilistic delay cell, which latches the first arriving rising edge, are shown in the upper half of the circuit. The cell consists of a transconductance stage, which behaves as a linear voltage-controlled current source, controlled by the voltage difference between the power supply and the function generator voltage. The power supply is $V_{power} = 10~V$, and the transconductance resistor is $R_{TC}$ = 4.9k$~\Omega$. This is followed by a programmable hysteresis stage to eliminate barrier-crossing errors. The hysteresis window is set by the ratio of the feedback resistor ($R_{F1}$ = 50k$~\Omega$) and the digital potentiometer ($R_{HTh1}$ = 1k$~\Omega$) and the input reference voltage ($V_{REF1}$ = 0.54~V). The digital output goes into a latching stage to catch and keep the first arriving rising edge. The reference path, shown in the bottom half of the circuit, consists of another noise removal hysteresis stage, with the feedback resistor $R_{F2}$ = 24.9k$~\Omega$, threshold resistor $R_{HTh2}$ = 100$~\Omega$ and $V_{REF2} = 5.5~V$. This is followed by a set of logic gates that enable the counter when the reference edge arrives and disables it when the latch is activated. Note that this enable signal is active low and is the length of the delay that we are interested in. The circled numbers in panel (b) correspond to the nodes in panel (a), from which experimentally measured time traces are collected and shown.  }
    \label{fig:ckt-diagram}
\end{figure*}

\section{Circuit Design and Experimental Setup}
\label{sec:CDES}
\subsection{\label{sec:circuit}Circuit design}


The circuit shown in Fig.~\ref{fig:ckt-diagram}(a) performs several key functions. First, it uses a programmable current source to bias the device into an initial, parallel state. Upon receiving an external start signal (the falling edge shown in Fig.~\ref{fig:ckt-diagram}(b)\circled{1}), the circuit applies a current step to the MTJ and subsequently measures the time before the MTJ switches to the antiparallel state. We measure this time digitally using an on-board counter and clock integrated circuit.

The details of the design are as follows. The circuit is composed of two separate paths that converge at logic circuits feeding a counter, namely the signal path and the reference path. The first stage of the signal path sets the current across the device using an operational-amplifier-based voltage-controlled current source with a fixed resistor ($R_{TC}$) and a programmable input voltage. The current is a linear function of the voltage and can be controlled to a precision of 0.2~{\textmu}A. The input to the voltage-controlled current source comes from a function generator, which applies the precise voltages required to operate the current source, and can be triggered by an externally controlled microprocessor to enable automation. 

The programmable current flows through the device, inducing random telegraph switching behavior, as illustrated in Fig.~\ref{fig:ckt-diagram}(b)\circled{3}. The voltage is then sent to a programmable hysteresis stage for thresholding and signal conditioning. Figure~\ref{fig:ckt-diagram}(b)\circled{4} shows the output of the programmable hysteresis stage that produces digital voltages, enabling direct integration with other digital circuit elements. To ensure the detection of the first rising transition, a set-reset latch is incorporated. This latch maintains a high output even after the SMTJ signal reverts to the parallel state, effectively serving as an indicator of a legitimate switching event. The output of the latch is shown in Fig.~\ref{fig:ckt-diagram}(b)\circled{5}.


In this configuration, the transition from the parallel to the antiparallel state results in the programmable hysteresis stage output rising from \texttt{0} to \texttt{1}. The speed of the amplifier utilized in the programmable hysteresis stage is a critical specification, as it dictates the maximum switching speeds that can be accurately captured.  Note that if the amplifier is too slow, it can cause short dwell times in the antiparallel state to be filtered by the circuit. This can prevent the detection of transitions to the antiparallel state that are short-lived and do not trigger the latching stage. In this work, the amplifier has a response time of $100$~ns, which is faster than the clock period of $500$~ns. Since the clock period determines the fastest times that can be measured, the speed of the amplifier is sufficient to capture transitions of interest. If the state switches back to the parallel state in less than a clock period, that transition would not be detected by this circuit. 
 
The transconductance stage, the programmable hysteresis stage, and the latching stage combine to form a single unit we call the probabilistic delay cell (PDC). The remaining circuits measure this delay and convert it to a digital value. \reviewing{The equations that govern the operation of the transconductance stage and hysteresis stage are described in appendix~\ref{Appendix: Equation}.}

The function generator that controls the input to the voltage-controlled current source also serves as an input to the reference path, which passes through independent threshold and logic circuits before reaching the counter to initiate the counting process. The role of the logic circuits is to define the size of the count window by enabling the counter immediately upon the occurrence of the triggering event (shown in Fig.~\ref{fig:ckt-diagram}(b)\circled{2}) and disabling it as soon as the set-reset latch is activated. This creates an active low pulse (shown in Fig.~\ref{fig:ckt-diagram}(b)\circled{6}), which has the same duration as our stochastic interval of interest. A critical design specification is to ensure that the signal path and the reference path are matched in terms of delay. This alignment is essential to prevent systematic timing errors in the count values introduced by the circuit components. Experimental measurements indicate a residual timing difference between the two paths of $625$~ns, which is a little larger than the $500$~ns period of the clock. This causes a systematic one-bit offset in count values generated by the counter. 

The programmable input voltage enables this circuit to accommodate devices with varying current biases. The tunable resistances in the programmable hysteresis circuits provide control over threshold hysteresis, facilitating effective operation with devices exhibiting different levels of noise activity. To ensure a broad temporal measurement range, two 8-bit counters are chained to create a 16-bit counter, while the clock rate can be externally adjusted over two orders of magnitude using a digital control word. All of the timing values in this paper are arrived at by multiplying the period of the clock by the value of the 16-bit counter. This circuit demonstrates an integrable approach to reading out analog temporal signals as well as an efficient approach to analog-to-digital conversion of stochastic temporal signals.

\subsection{\label{sec:setup}Experimental Setup}


The experimental setup consists of the aforementioned circuit on a printed circuit board, mounted onto test fixtures on a probe station for mechanical stability. High-frequency ground-signal (GS) probes are used to make contact with the SMTJs mounted on the probe station stage. The SMTJs exhibit a magnetic easy-axis out of the plane of the device through the following material stack structure:  Si base / SiO2 / TaN / [Co(0.5)/Pt(0.2)]6 / Ru(0.8)/ [Co(0.6)/Pt(0.2)]3 / Ta(0.2) / Co(0.9) / W(0.25) / CoFeB(1) / MgO(0.8) / CoFeB(1.4) / W(0.3) / CoFeB(0.5) / MgO(0.75) / Ta(150) / Ru(8). The numbers in parentheses refer to layer thicknesses in nanometers, while the numbers beside square brackets show the bilayer repetitions. These devices, circular in shape, have approximate diameters of 150~nm. They exhibit a tunneling magnetoresistance (TMR) near $120~\%$ at room temperature and possess a resistance-area product of 10$~\Omega$~µm$^2$. Breakdown of the MgO tunneling barrier has been observed in these devices at voltages around 0.7~V (corresponding to currents above roughly 1~mA).

The experiment is designed for full automation, enabled by an external trigger. The setup involves connecting the circuit board to a microcontroller, which interfaces with a computer. The process begins when the computer sends a start signal to the microcontroller, prompting it to program and trigger the function generator. The function generator then initiates the current step, setting the signals in motion through the circuit, ultimately resulting in a count value at the output pins of the 16-bit counter. The microcontroller reads this value and transmits it back to the computer, where it is decoded and stored for further analysis. This automated loop facilitates repeated sampling, allowing the experiment to be scaled and repeated as needed. Automation enables efficient data collection, facilitating robust analysis of stochastic delay elements under varying conditions.

\section{\label{sec:delay_char}Properties of probabilistic delay cells}


\begin{figure}
    \includegraphics[width=8.5cm]{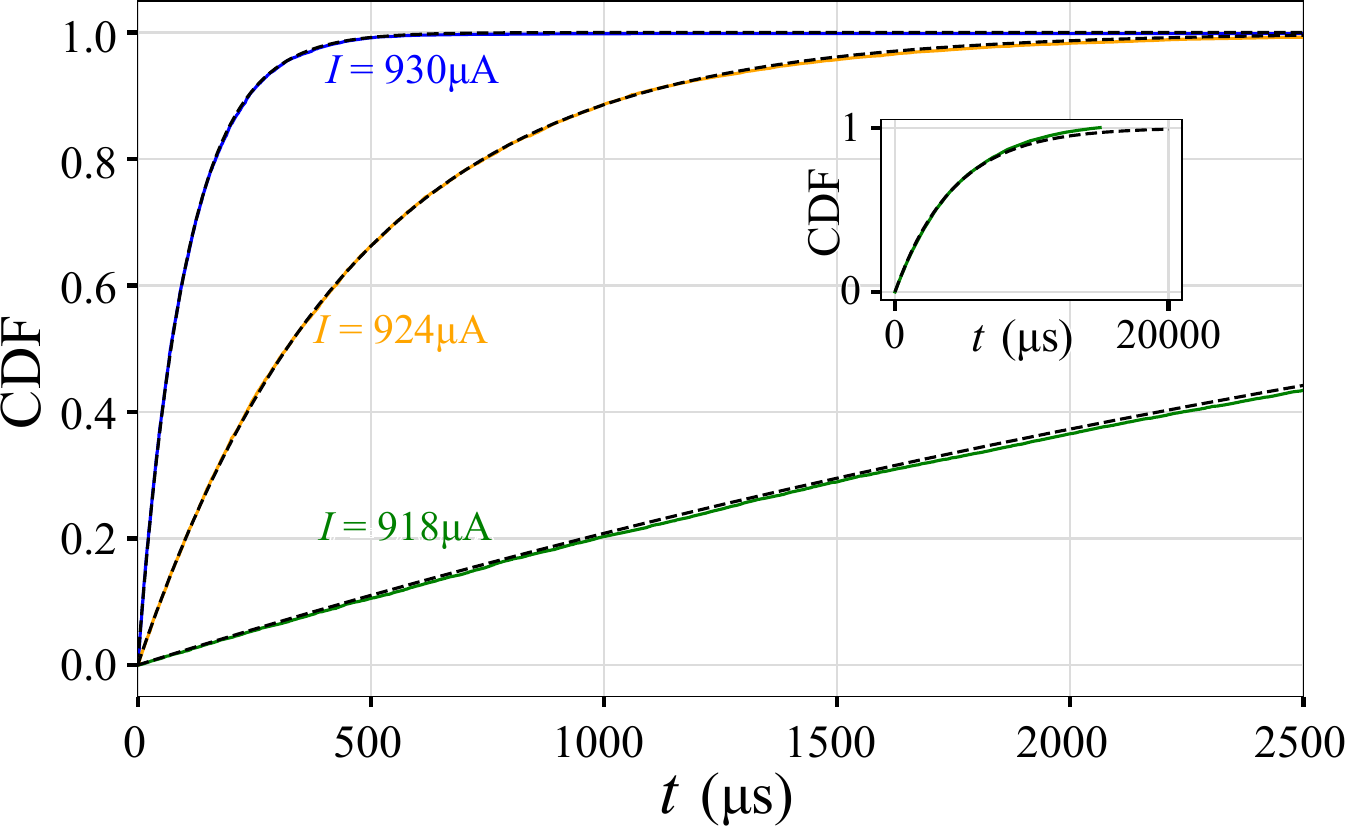}
    \caption{Cumulative distribution function (CDF) of switching times: For three current levels (918~µA, 924~µA, 930~µA), the CDF demonstrates the tunable nature of probabilistic delay cells. Higher currents result in faster switching times, and lower currents give longer switching times, illustrating the input current control of the timing distribution of the device. The colored lines represent the data, while the dotted lines represent the theoretical fits to Eq.~\ref{eq:cdf}. The inset shows the same data as the lowest current level while showing the long-time behavior of the CDF. The deviation from the fit at long time scales is a result of device variability.   }
    \label{fig:statistics}
\end{figure}

We model the switching behavior of the device as a Kramer's escape problem, with the statistics of magnetization reversal being described as a Poisson process. This results in an exponential distribution of delay times between switching events, parameterized by a rate $\lambda$. The inverse of this rate is the mean switching time that characterizes the probabilistic delay cell and is denoted by $\tau$, which can be extracted from the dwell time distribution shown in Fig.~\ref{fig:intro}(c). The cumulative distribution function (CDF) of the exponential distribution that describes the behavior of the delay cell is 
\begin{align}
\label{eq:cdf}
    F(t) = 1-e^{-\lambda t}
\end{align}
The CDF for different values of current and their fits to Eq.~\ref{eq:cdf} are plotted in Fig.~\ref{fig:statistics}.
An important characteristic of the probabilistic delay cell is its tunability, which allows for precise control over the distribution of delay times. By adjusting the amplitude of the current step applied to the device, the mean of the distribution can be adjusted with larger current steps shortening the delay times, and smaller current steps resulting in longer delay times.  

\reviewing{Though a model with a single rate parameter fits the experimental data as shown in Fig~\ref{fig:statistics}, this may not always be the case---device sizes and operational switching time scales will likely decrease as the technology matures. There are several effects that might contribute to deviations from the observed exponential behavior when scaling to smaller and faster technologies, including transient temperature variations due to the current step and deviations from the two-state description implicit in \eqref{eq:cdf}. A macrospin model with continuous degrees of freedom is the simplest model to show deviations from a two-state model. We analyze an easy-axis macrospin system as a representative model, and consider the evolution of the distribution when a bias term pinning the system to the P state is instantaneously switched to a bias favoring the AP state. Using an appraoch related to that in Ref.~\cite{butler2012switching}, we solve the time-dependent Fokker-Planck equation under AP bias starting with the equilibrium distribution for a bias favoring the P state. At long times, the probability of the particle being in the AP state increases as in \eqref{eq:cdf}, but at short times, there are corrections to this behavior. While further study is warranted, we find that these corrections are substantial for low-barrier MTJs but can be made unimportant for large enough barriers ($\Delta E/ kT > 15$).

This analysis agrees with previous studies on the application of current steps on magnetization reversal dyanmics~\cite{butler2012switching,mcgoldrick2022settling}.  Micromagnetic and macrospin simulations~\cite{mcgoldrick2022settling} show that micromagnetic response to a current step not captured by macrospin models depends on additional device properties such as diameter and damping factor. For the systems they studied, they found that the time scales of these effects were between 100~ps and 1~ns. Autocorrelation studies can be conducted on devices to determine how long it takes for these fast transients to dissipate. In the long-time limit, these effects are small and difficult to observe at the time scales at which our experiment is run, but they do impose an upper limit on the speed at which designs based on these circuits can accurately sample exponential distributions.}

\begin{figure}
    \includegraphics[width=8.5cm]{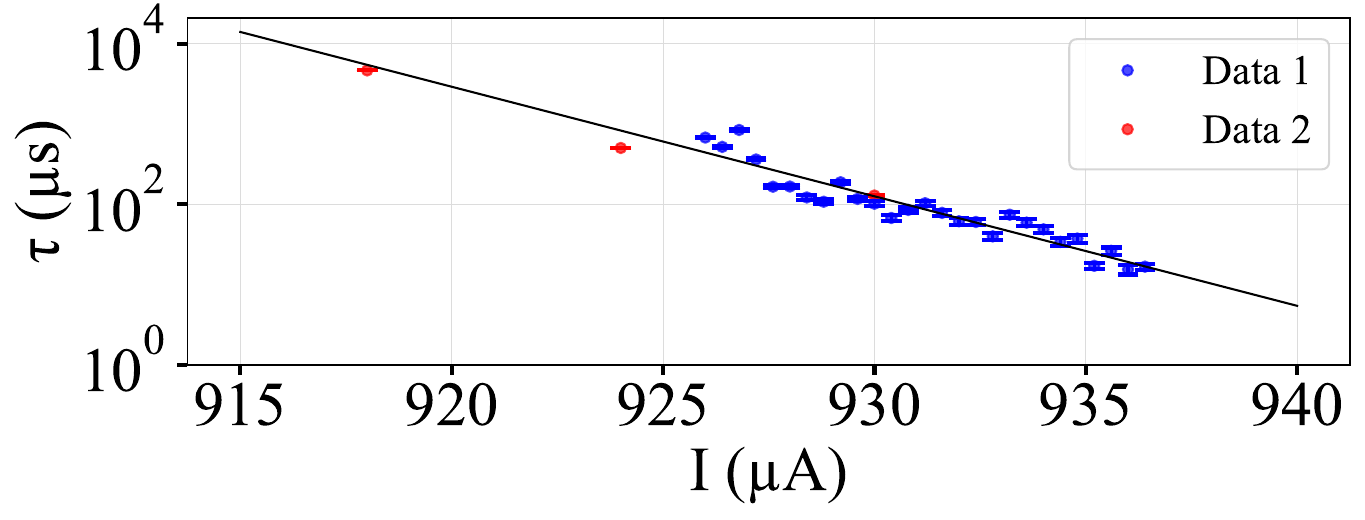}
    \caption{Exponential dependence of mean switching times with current: The figure shows the exponential relationship between current $I$ and mean switching time \( \tau \), as in Eq.~\eqref{eq:tau_fit}. Both datasets are collected on the same device on different days with different current ranges. The error bars on the experimental data show single standard deviation uncertainties in the mean. }
    \label{fig:meanswitch}
\end{figure}

The mean value of the distribution depends exponentially on the current; we plot experimental observations of this value in Fig~\ref{fig:meanswitch}. \reviewing{In a two-state model, the relationship between the switching time and the current flowing through the device is
\begin{align}
\label{eq:tau_fit}
 \tau_{\text{P}}(I) = \tau_0 \exp\left[\frac{\Delta E}{k T} \left(1 + \frac{I}{I_c}\right)^\alpha\right]   
\end{align}
where \( \tau_{\text{P}}(I) \) is the mean dwell time in the parallel state, \( \tau_0 \) is a characteristic time, \( \Delta E \) is the energy barrier, \( k \) is the Boltzmann constant, \( T \) is the temperature, \( I \) is the applied current, \( I_c \) is the critical current, and we take $\alpha=1$ \footnote{Different models predict different values for the exponent $\alpha$~\cite{sun2022metrology}. While our results are consistent with $\alpha=1$, they are not inconsistent with other values because the fractional variation of the current in our measurements is so small. Taking any value for $1\le\alpha\le 2$ does not give fit results that are more sensible physically.}.
While this equation provides a functional form that describes the experimentally observed behavior, the values extracted from fitting the equation to the data do not give physically meaningful values of $\tau_0$, $\Delta E$, and $I_c$.}  The wide range of delay tunability, almost two orders of magnitude, is useful when the timescale of the device operation needs to be matched with the timescale of other aspects of the circuit. Specifically, we will see this in Sec.~\ref{fig:application}, where such tunability enables crucial functionality when designing a Metropolis-Hastings stepper or a weighted random sampler.

One drawback of the devices measured in this work is that the device properties appear to drift. In Fig.~\ref{fig:meanswitch} the data points differ from the exponential fit by more than expected from the statistical uncertainties. We interpret these excessive deviations as due to such drift. More than ten thousand switching events were collected at each current level to reduce the statistical uncertainty in each of the extracted values. Using those uncertainties, the reduced chi-squared statistic for the fit is $\chi^2$ = 10.68, which is significantly larger than 1, indicating that there is variation in the behavior that is not captured by the statistical uncertainty. ~\reviewing{To study this drift, we measured the stability of the mean dwell time of the anti-parallel state at a constant current. The result of this experiment, shown in Appendix~\ref{Appendix: Drift}, shows a statistically-significant drift in device properties. We conclude that this set of devices is not technologically mature enough to generate stable distributions. Such behavior was also reported with the same device stack in previous work~\cite{gibeault2024programmable}. We do not have an explanation for the drift. One possibility is fluctuations in ambient temperature but the timescale of the fluctuations in device properties reported in Appendix~\ref{Appendix: Drift} seems fast for large fluctuations in ambient temperature. This can result in a single current level manifesting a variety of exponential distributions with slightly different rate parameters, depending on the present state of drift.} This behavior can also distort the results when measuring curves for long mean dwell times because the measured behavior is then a combination of results with multiple exponentials. This deviation from exponential behavior could be an explanation for the deviation between the fit and the data seen in the inset of Fig.~\ref{fig:statistics}. Characterizing the physical origin of this drift in device properties is beyond the scope of this work. 

\section{\label{sec:appliction}Applications}

Sampling from specific distributions of interest in hardware is an important problem~\cite{thomas2009comparison} and can benefit from acceleration as various machine learning and statistical computational models become more prevalent. In conventional digital computing systems,  uniform distributions are commonly used as a general source of randomness; Gaussian distributions are used for discrete-time simulations; and exponential distributions are used for discrete event simulations~\cite{thomas2009comparison}.  Both conventional computing systems~\cite{jun1999intel,NVIDIATRNG} as well as recently developed probabilistic computing systems~\cite{singh2024cmos,niazi2024training} rely on evenly Bernoulli distributed ($p=0.5$) random bits provided by their representative hardware random number generators. Sampling from other arbitrary distributions of interest is performed using alias tables, which were proposed by Walker~\cite{walker1977efficient}. Alias tables, which are effectively weighted look-up tables, enable conversion between probability distributions and have been implemented in graphical processing units (GPUs)~\cite{wang2021skywalker} and FPGAs~\cite{thomas2008sampling}. The prevalent use of alias tables points to the potential efficiency gains to be found in developing efficient weighted random samplers that can be used to directly sample from arbitrary distributions of interest.  

\reviewing{We compare our approach with other spintronic approaches, which are typically based on measuring the instantaneous state of a stochastically changing MTJ. A comparison against all modern approaches to random number generation is beyond the scope of this work. We aim to convince the reader that exponentially distributed random number generation has the potential to become energy and latency-competitive with Bernoulli-distributed generators.} 

\begin{figure}
    \includegraphics[width=8.5cm]{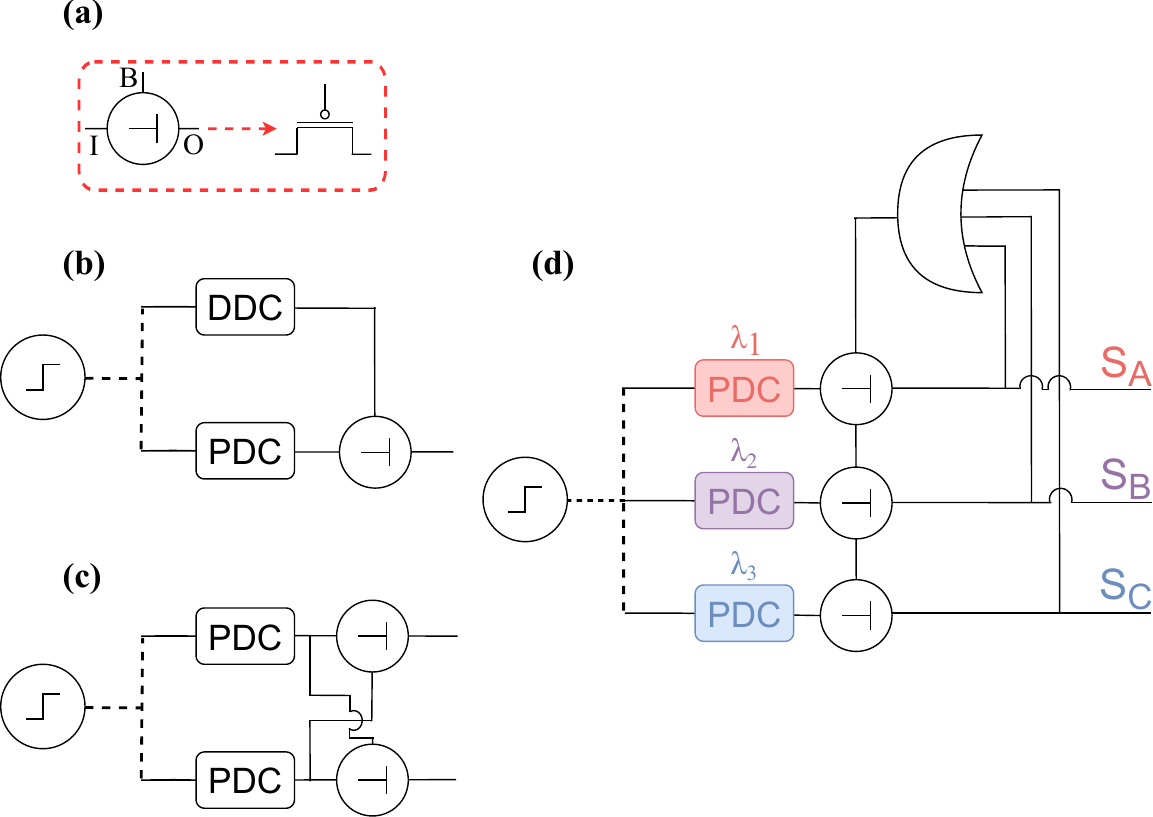}
    \caption{Probabilistic sampling with temporal circuits: Panel (a) shows an inhibit gate, a primitive of temporal computing that can used to block rising edge inputs from propagating through them, based on the relative arrival time between two input signals. Input I is the inhibited input, while input B is the blocking input. If input B arrives first, input I is blocked, and output O remains fixed at \texttt{0}. If input I arrives first, it is allowed to pass through, causing output O to rise to \texttt{1}. Panel (a) also shows a single transistor implementation of the inhibit gate from~\cite{tzimpragos2019boosted}. Panel (b) shows the simplest version of a circuit that can make Boolean decisions based on a temporal sample from an exponential distribution. A deterministic delay cell (DDC) and a probabilistic delay cell (PDC) are fed into the B (for blocking) and I (for inhibited) inputs of the inhibit gate, shown in panel (b). The output (O) transitions to a Boolean true value if the probabilistic delay cell fires earlier than a deterministic delay value. Panel (c) extends the notion in panel (b) with two cross-coupled probabilistic delay cells that inhibit each other based on the first arriving probabilistic event. This can be used to generate a biased coin flip when the rates of the two probabilistic delay cells are different from each other. Panel (d) extends the two-input circuit from panel (c) into an $n$-input version, where the OR gate selects the first arriving input and blocks the other inputs such that only the first arriving signal makes it to its output. The final one-hot word, determined by the wire (one of $S_A, S_B, S_C$) that transitioned to \texttt{1}, encodes the result of the random draw. The probability that output $j$ fires first is described by equation~\ref{eq:exponential_clocks}.}
    \label{fig:application}
\end{figure}

In temporal computing~\cite{advait2014race} information is encoded in the \reviewing{analog} arrival time of rising edges (transitions from a low voltage to a high voltage). Decisions can be made by racing signals against one another, with the earlier arriving signals blocking the later arriving ones. The blocking operation is performed by the inhibit gate as shown in Fig.~\ref{fig:application}(a). If the inhibiting signal (represented by input B for ``blocking'' in Fig.~\ref{fig:application}(a)) arrives first, the signal labeled ``inhibited'' (represented by input I in Fig.~\ref{fig:application}(a)) gets blocked (inhibited), resulting in a \texttt{0} output. If the input signal arrives first, it passes through the inhibit gate, causing the output to rise to \texttt{1} (\textit{i.e.} it is not inhibited). In its simplest form, an inhibit gate can be implemented with a single $p$-type metal oxide semiconductor (PMOS) transistor~\cite{tzimpragos2019boosted}. When a rising edge arrives at the gate of the PMOS transistor, it turns the transistor off, closing the gate and maintaining the existing voltage level by preventing additional charge from coming in or trapped charge from leaking. 

Fig.~\ref{fig:application}(b), utilizes the inhibit gate shown in Fig.~\ref{fig:application}(a) in its simplest possible configuration, with a deterministic delay cell (DDC) feeding into its blocking (B) and a probabilistic delay cell (PDC) feeding into its inhibited (I) inputs respectively. \reviewing{As described in section~\ref{sec:delay_char}, the probabilistic delay cell generates an exponentially distributed analog delay time.} The circuit operates by allowing all probabilistic samples with a value smaller than the deterministic analog delay value to pass through while blocking all edges that arrive later. Hence, this circuit implementation encodes the probability that a sample drawn from an exponential distribution has a value smaller than a fixed delay value $T_D$, which is the delay of the deterministic delay cell. This probability is equal to the value of the cumulative distribution function of the probabilistic delay cell at time $t=T_D$. Various complementary metal oxide semiconductor (CMOS) implementations of deterministic delay cells have been proposed in the literature~\cite{JieGuTdImage,sayal201912,madhavan20174,kim20149}.

This circuit has a few potential uses. One is to generate Bernoulli-distributed random bits, like those generated by sampling the state of an MTJ. For this application, the median time of the probabilistic delay cell can be determined by evaluating the inverse CDF at $1/2$. Half of the events occur before $\tau\ln(2)$, while the other half of the events occur after. Setting the delay of the deterministic delay cell $T_D$ equal to this value in the circuit shown in Fig.~\ref{fig:application}(b), causes the inhibit gate to block the output of the probabilistic delay cell with a probability of $1/2$. By choosing $T_D = -\tau\ln(1-p)$, one can bias the distribution to have probability $p$ instead of $1/2$. For this application, the temporal cell is unlikely to have a significant advantage over sampling the state of the device. Advantages occur in applications that use the sampling of the native exponential distribution beyond generating Bernoulli samples. A discussion comparing the two approaches can be found in Sec.~\ref{sec:discussion} 

Such a probabilistic delay circuit can be used in a temporal version of probabilistic computing applications. One example is a stochastic search using a Markov chain Monte Carlo sampler with Metropolis-Hastings transition rules. This approach attempts to generate a thermal distribution of states by stepping through configuration space with probabilities that depend on the energy change due to the step compared to the temperature of the simulation. In the simplest serial version of such an approach, the configuration space is searched by randomly flipping spins and evaluating the change in energy ($\Delta E_i$) due to a single spin flip indexed by $i$. If this value is negative, then the flip is accepted, but if the value is positive, the flip is accepted with probability equal to the ratio of the proposed and original Boltzmann-distributed probabilities. This approach is common in large statistical simulations~\cite{kirkpatrick1983optimization,aadit2022massively} and can benefit from the acceleration provided by the circuits described in Fig.~\ref{fig:application}(b).  

As shown in Fig.~\ref{fig:application}(b), two temporal signals are input to the inhibit gate in the circuit with values $t_0 + \Delta t$ and $t_0 +\Delta E_i/W$, where $W$ is some characteristic power that controls the granularity of the embedding into the temporal domain, and $\Delta t$ is an exponentially distributed random variable with rate parameter $W\beta$. The $\Delta E_i/W$ value is encoded in the deterministic delay cell while the $\Delta t$ is encoded in the probabilistic delay cell. The deterministic signal then uses the circuit in Fig.~\ref{fig:application}(b) to inhibit the stochastic signal; that is, if $t_0 + \Delta E_i/W < t_0 + \Delta t$, the output of the inhibit ($\dashv$) gate will be set to Boolean false; otherwise, it will be Boolean true. The probability that the inhibit gate outputs false is
\begin{equation}
    P(\Delta t < \Delta E_i/W) = 1 -\exp\left[-\Theta(\beta \Delta E_i)\right],
\end{equation}
from the cumulative distribution describing the $\Delta t$, where $\Theta$ is the Heaviside theta function; this step function appears because the probability that $\Delta t < 0$ is zero. We then invert this Boolean output to produce a final Boolean value which is true with probability $e^{-\Theta(\beta\Delta E_i)}$. The value of this random bit tells us precisely whether or not to accept the proposed transition under the Metropolis-Hastings algorithm. In the case of an Ising model, we could use this signal directly to update a stored Boolean value of some associated spin $s_i$. We leave the details of constructing $\Delta E$ and coupling spins into an Ising machine for future work.

Sampling from exponential distributions forms a key building block of a more general kind of sampling from arbitrary distributions, known as weighted random sampling. Weighted random sampling involves picking from a discrete set of items indexed by $j$, with a weight $w_j$. This fundamental computational task has a wide variety of applications in various fields, such as machine learning~\cite{byrd2019effect}, statistical simulation~\cite{hubschle2022parallel}, and database systems~\cite{kurant2011walking,kalal2008weighted}. Conventionally, these methods rely on uniform random samplers and use algorithmic methods to generate weighted samples. In this work, we rely on the principle of the ``exponential clocks''~\cite{hubschle2022parallel} method in order to generate weighted random samples with probabilistic delay cells and temporal computing primitives. 

The exponential clocks method described in~\cite{hubschle2022parallel,arratia2013amount} asks: given $n$ exponentially distributed clocks with various rate parameters $\lambda_j$, what is the probability $P_j$ that the $j^{th}$ clock will ring first? The solution is given by 
\begin{equation}
\label{eq:exponential_clocks}
    P_j = \frac{\lambda_j}{\lambda_1 + \lambda_2 ... + \lambda_n} .
\end{equation}
Such a scheme can be naturally implemented with temporal operators and exponentially distributed probabilistic delay cells with tunable rates. The simplest case is $n=2$, as shown in figure \ref{fig:application}(c), with two probabilistic delay cells feeding into two cross-coupled inhibit gates. The first-arriving of the two delay signals reaches the output before the other signal, similar to the previous case, and blocks the other one from reaching the output. The output that transitions to a Boolean true is the current sampled index. 

This can be scaled up to $n$ elements, as shown in Fig.~\ref{fig:application}(d), by using an OR gate to select the first arriving input signals. The output of the OR gate feeds the inhibiting inputs (input B) of all of the signals, allowing only the first arriving signal to pass through, while effectively blocking all other signals from reaching the output. The output that transitions to a Boolean true represents the index of the sampled element. This is equivalent to rolling an $n$-sided die with arbitrary (but normalized) probabilities for each side, where the biases are dictated by the ratio of the rate parameters of the different SMTJs as shown in equation~\ref{eq:exponential_clocks}. Note that since the rate parameters in Eq.~\eqref{eq:exponential_clocks} are themselves exponential functions of the current, current-addressing the devices gives direct access to a Boltzmann distribution $P(j) \propto e^{-I_j/I_\beta}$ where $I_\beta$ is a characteristic current.

\section{\label{sec:discussion}Discussion}

In this work, we demonstrate how concepts from temporal computing and probabilistic computing literature can reinforce each other, resulting in techniques that can benefit both computing paradigms. First, we use the flagship device technology of probabilistic computing, namely the superparamagnetic MTJ, to create a novel primitive for temporal computing: the probabilistic delay cell with tunable mean delay $\tau$, or equivalently, the rate parameter $\lambda$. Conventionally, only deterministic delay cells have been used in temporal computing, limiting the ability of temporal systems to perform probabilistic computations. Replacing the delay element in temporal memories as described in ~\cite{advstatemachine} with a probabilistic delay cell opens up an array of statistical operations that could not be previously implemented in the temporal domain. Such operations are particularly useful in the context of energy-efficient architectures for machine learning operations such as convolution~\cite{Rhys}, matrix multiplication~\cite{sayal201912,nair2021microarchitecture}, median filtering~\cite{JieGuTdImage} and others.

We also show how using the time domain for sensing and computation allows for the sampling of programmable exponential distributions, which are common to energy-based probabilistic models of computing. Though the exponential dwell time distributions of SMTJ devices are well understood and often used in the characterization of device properties~\cite{hayakawa2021nanosecond,schnitzspan2023nanosecond,kaiser2022hardware}, they have not been directly used in sampling or computation. The temporal information~\cite{advait2014race} representation coupled with simple CMOS primitives allow readout and energy-efficient translation of exponentially distributed probabilistic delay times into digital values. These values can be used to flip spins in an Ising machine~\cite{aadit2021computing} or change the neuron state in a Boltzmann machine~\cite{niazi2024training}, both of which can utilize exponential distributions to set their state acceptance probabilities. 

The exponential behavior that results from barrier crossing, such as the one proposed in this work, is also seen in other device technologies. Single-photon-avalanche-diode(SPAD)-based circuits, which also generate Poissonian event statistics derived from statistics of photon arrival times, have been recently proposed as another CMOS compatible technology to solve Ising and Potts models ~\cite{whitehead2023cmos}.  These circuits have also been used with a clock period tuned to the median time of switching events to generate uniform random numbers in a temporal way, similar to the approach presented in this work~\cite{stanco2020efficient}. Metal oxide memristors, another back-end-of-the-line compatible technology, also exhibit random telegraph noise in their high resistance state~\cite{choi2014random,li2021random}. This is attributed to the redistribution of oxygen vacancies in the device, which can also be used as another source of generating exponentially distributed switching events, but lacks the advantage of current tunability.

Finally, some energetic considerations remain regarding the integrability and efficiency of circuits presented in this work. In the composite circuit described in Fig.~\ref{fig:ckt-diagram}(a), all subcircuits can be efficiently incorporated into integrated implementations except for the timing measurement circuits, which will incur significant energy costs~\cite{advait2014race} if repeatedly instantiated with each delay cell. It has been previously shown that the energy cost of such timing measurements can be amortized by using them only at the final stages of a temporal computation~\cite{madhavan20174,JieGuTdImage}. On the other hand, the functionality provided by the transconductance and hysteresis stages on a printed circuit board can be efficiently performed by more compact integrated implementations. The voltage-controlled current source, shown in Fig.~\ref{fig:ckt-diagram}(a), can be replaced by a self-resetting cascode stage~\cite{narayanan1996static}. The cascode stage can provide the functionality of the current source, while the self-resetting design of the circuit allows for the current to be stopped as soon as the SR latch switches to a high state, hence further reducing static power consumption. Similarly, CMOS hysteresis comparators (also known as Schmitt triggers~\cite{filanovsky1994cmos}) and latches are relatively cheap to implement, requiring only a handful of transistors. \reviewing{Assuming optimistic MTJ properties, such as a 1~V power supply, a 10~{\textmu}A operating current, and a 100~ns operation time, the energy cost of each operation can be estimated as being around 1~pJ per switch. This value is a few orders of magnitude above CMOS gate transitions, which range from tens to hundreds of femtojoules. Hence, even for optimistic device properties, the energy consumption in the SMTJ itself dominates over that in the integrated CMOS circuitry. If it is possible to fabricate MTJs with nanosecond scale time scale fluctuations, state sampling techniques for such MTJs would be more energy-efficient per sampled bit. However, in applications that would require many Bernoulli trials in order to sample an exponential variate, this temporal approach would still be advantageous. This is discussed next.

Both the instantaneous state-encoded and temporally-encoded approaches have advantages and disadvantages when it comes to generating random samples. Some metrics that can enable such a comparison are the number of random bits per sample, samples per unit of time, and the number of samples per unit of energy. The values of these metrics are different for both approaches and hence, a different set of device properties are optimal for each approach. 

Nanosecond switching SMTJs are suitable for the state-encoded approach since they are able to produce a large number of random bits per unit time. As the operating currents of these devices are lowered, they can lead to very energy-efficient readouts. Assuming a 10~$\mu$A operating current at a volt, and a sub 5~ns autocorrelation time, such an approach can lead to $2\times10^8$ random samples per second, with at most one bit per random sample, and an energy cost of 50~fJ per switching event in the device. Since these devices have small barriers, they are more prone to transient timing fluctuations (as discussed in Section~\ref{sec:delay_char}), making them unsuitable for the temporal approach. 

Medium barrier magnets are more suited to the temporal approach since they can produce reliable exponential distributions, with transient effects playing a smaller role in their switching behavior. More importantly, such an approach produces an analog temporal quantity, which contains more than one bit of information. How such a temporal quantity is quantized has important implications. Assuming similar device operating ranges such as a 10~$\mu$A operating current at a volt, and a 100~ns operating window partitioned into 8 time-bins encoding a 3-bit binary value, such an approach has the potential to produce $3\times10^7$ random samples per second with an energy cost of 1~pJ per transition, or 333~fJ per bit. This number can be further reduced by pulsing the device with a larger current leading to faster switching, or by increasing the number of bits that can be extracted per switching event. The degree of variation in device properties and the precision of temporal supporting circuits will ultimately decide the number of bits per sample.

Given the instability in device behavior, both approaches need significant improvement in manufacturability within proper device margins to generate uniformly distributed bits. Assuming an optimistic scenario where manufacturable margins are achieved, both approaches require finely tuned current pulses to achieve the $p=0.5$ case. That being said, variations in device parameters affect the temporal approach and the instantaneous-state encoded approach differently. While variations in the temperature or the energy barrier can cause the devices to operate faster or slower, to first order, they do not impact the relative dwell time distributions for a device biased at $I_{50-50}$. On the other hand, the temporal approach is much more sensitive to such variations since they can have an exponential effect on the measured time. Post-fabrication tuning becomes important to mitigate such issues. This suggests that such an approach is more suitable for temperature-regulated environments of data centers rather than edge applications with wide temperature ranges. 

Once a set of devices has been fabricated, in order to improve the number of random samples per unit time, the state-encoded approach needs to change device properties such that the device operates at a faster time scale for the same $I_{50-50}$ current. This cannot be done without fabricating a new set of devices. On the other hand, the speed at which the temporal approach operates can be tuned by adjusting the size of the deterministic delay and the current pulse, both of which can be performed post-fabrication. Since tuning the size of the current step can lead to an exponential reduction in the delay of the probabilistic delay cell with a linear increase in current, this has the potential to save a lot of energy. Though limited by the theoretical considerations described in Sec.~\ref{sec:delay_char}, the temporal approach allows for more post-fabrication flexibility in operation when compared to the state-encoded approach and a potentially faster operation with slower devices. 

At the application level, when interested in Bernoulli events that are exponential functions of some quantity, such as in the metropolis hastings algorithm, the state-encoded approach requires numerous Bernoulli events to effectively approximate such a distribution~\cite{dughmi2021bernoulli}. In such a case, each Bernoulli event is sampled with the instantaneous-state-based approach requiring an SMTJ switching event. \emph{On the other hand, in the temporal case, a single switching event of the SMTJ naturally produces an exponentially distributed analog value, which can be directly used in temporal computing circuits.} The speed of this transition can also be controlled by changing the current flowing through the device, allowing for the energy to be further reduced with a similar tradeoff as described above. For generating exponentially distributed random variates, therefore, the temporal approach is likely faster and more efficient than state-encoded methods. In this way, the circuits described in this work pave the way for integrated implementations of hybrid-probabilistic-temporal computers.
}

\section*{Acknowledgements}
TA, SG, DPL, and AM acknowledge support under National Science Foundation grant No.~CCF-CISE-ANR-FET-2121957. AM acknowledges support under the National Institute of Standards and Technology Cooperative Research Agreement Award No.~70NANB14H209 through the University of Maryland. PT and UE acknowledge support under the Agence Nationale de la Recherche StochNet Project award No.~ANR-21-CE94-0002-01. The authors acknowledge J. Langer and J. Wrona from Singulus Technologies for the MTJ stack deposition and N. Lamard, R. Sousa, L. Prejbeanu, and the Upstream Technological platform PTA, Grenoble, France for the device nanofabrication. The authors thank A. Hakam, B. Zink, and B. Rippard for helpful discussion and feedback. 

\appendix
\reviewing{

\section{Measurement of drift in device state}
\label{Appendix: Drift}

This section describes in more detail the drift in the device state reported in the paper. A static current is applied that nominally keeps the device close to the fifty-fifty state. The device's output is then continuously recorded over 200 s, collecting more than 60,000 switching events, after which the data is divided into 30 successive time windows. Each data point represents the average of more than 2000 transitions within each window. The probability of being in the antiparallel state is calculated for each bin and plotted in Fig.~\ref{fig:appendix1}. The large sample populations at each point on the curve provide a level of certainty that the time-dependence of the curve represents an underlying instability in the device rather than statistical fluctuations around a stable probability. In a newer batch of devices, we have noticed that an external magnetic field serves to reduce the instability, but those devices have different device stacks and material properties, and hence operating current ranges. Discussion of such a magnetic-field-based stabilization of these devices is beyond the scope of this work. 
\begin{figure}
    \includegraphics[width=8.5cm]{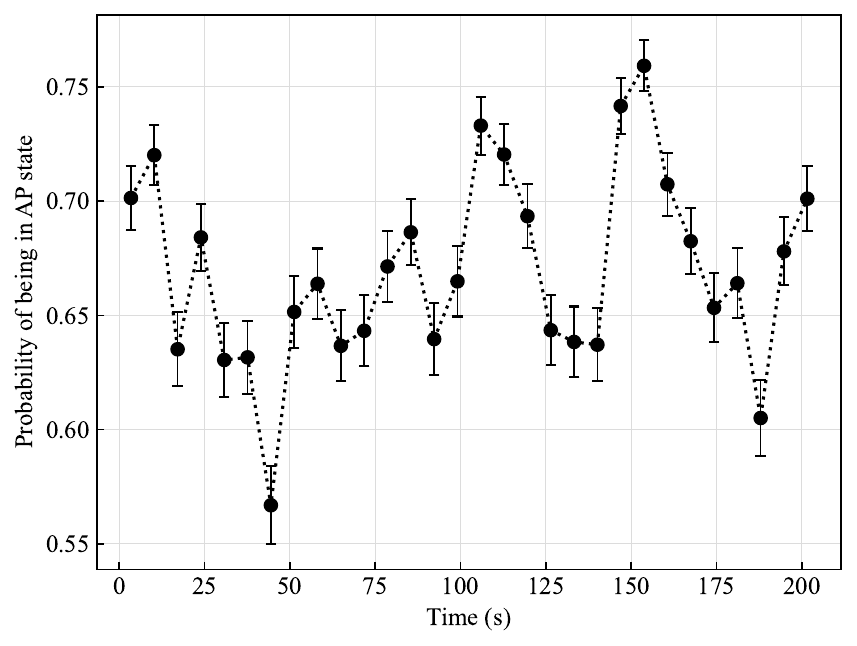}
    \caption{\reviewing{The probability of the device being in the AP state for a fixed current analyzed for successive time windows. With a fixed constant current, the mean dwell time in the AP state is monitored over 205~s and computed separately in thirty bins. There are approximately 2000 dwell events in the AP state for each bin. The error bars indicate single standard deviation uncertainties in the mean in each bin. Overall 30 bins, the average uncertainty in the mean is 0.015~s. If the device properties remained constant between bins, the width of the distribution of mean values should be close to the average uncertainty of the means over the bins. Instead, it is 0.042~s, almost three times larger, indicating that some property of the device or the measurement setup is drifting on the time scale of the bin size.}}
    \label{fig:appendix1}
\end{figure}
\section{Circuit equations for amplifier stages}
\label{Appendix: Equation}
\subsection{Transconductance Stage}
This section describes the equation for the transconductance stage in Fig.~\ref{fig:ckt-diagram}(a). The transconductance stage acts as a tunable current source for the SMTJ by using the high gain of the operational amplifier to pin the voltages across the resistor $R_{TC}$. This causes the current through the device to be solely determined by the resistance value, and is given by
\begin{equation}
I_{SMTJ} = \frac{V_{R_{TC}}}{R_{TC}} = \frac{V_{Power} - V_{in}}{R_{TC}} ,
\end{equation}
where $I_{SMTJ}$ is the current flowing through the device, $V_{power}$ is the power supply voltage and $V_{in}$ is the input voltage applied to the positive terminal of the amplifier. 

\subsection{Hysteresis State}
This section describes the equation governing the hysteresis stage presented in Fig.~\ref{fig:ckt-diagram}(a). $V^+$ is the voltage at the non-inverting terminal of the operation amplifier which is connected to the output node $V_0$  via the feedback resistor $R_F$ and to the reference node $V_{REF}$ via the resistor $R_{Hth}$. The inverting input node $V^-$ is connected to the input signals for each stage. The voltage at node $V^+$ is given as
\begin{equation}
V^+ = \frac{R_{HTh}}{R_{HTh} + R_F} \cdot V_0 + \frac{R_F}{R_{HTh} + R_F} \cdot V_{REF} .\end{equation} 
The output voltage $V_0$ swings between the supply rails (i.e. $V_{DD}$ and ground), effectively driving the non-inverting voltage to the higher and lower threshold values denoted by $V_{TH}$ and $V_{TL}$ respectively. 

The dynamics of the circuit are described as follows. If we start with the condition that $V^- > V^+$, then the output $V_0$ starts at the lower rail---ground, in our design---with $V^+$ $=$ $V_{TL}$. When $V^-$ transitions such that $V^- < V^+$ (lower threshold crossing), $V_0$  saturates to $V_{DD}$, and $V^+$ transitions to $V_{TH}$. In this new state, $V^- < V^+$ and $V_0 = V_{DD}$. When $V^-$ transitions such that \(V^- > V^+\)(higher threshold crossing), $V_0$  swings back to ground. The equation of the higher and lower threshold voltages for node $V^+$ is now given as
\begin{equation}
V_{TH} = \frac{R_{HTh}}{R_{HTh} + R_F} \cdot V_{DD} + \frac{R_F}{R_{HTh} + R_F} \cdot V_{REF} , \end{equation}
and
\begin{equation}
V_{TL} = \frac{R_F}{R_{HTh} + R_F} \cdot V_{REF} , \end{equation}
where $V_{TH}$ is the higher threshold voltage, and $V_{TL}$ is the lower threshold voltage.
 }






\bibliography{references}
\end{document}